
\magnification=1200
\def\sect#1{\vskip 8mm \noi {\bf #1} \vskip 6mm}
\def\section#1{\vskip 8mm \noi {\bf #1} \vskip 6mm}
\def\subsection#1{\vskip 6mm \noi {\bf #1} \vskip 4mm}

\def\subsub#1{\vskip 4mm \noi {\it #1} \vskip 4mm}
\def\subsubsection#1{\vskip 4mm \noi {\it #1} \vskip 4mm}
\def\Fig#1{\vskip 2mm \cen {-- Fig. #1 --} \vskip 2mm}
\def\Ha{H$_\alpha$}

\def\Mgal{\times 10^{11}\Msun}
\def\v{\vskip 2mm}

\def\noi{\noindent}

\def\title#1{\centerline {\bf #1}\vskip 8mm}
\def\author#1{\centerline{#1}\vskip 2mm}
\def\inst#1{\centerline{\it #1}\vskip 4mm}
\def\abstract#1{{\centerline{\bf Abstract}}\vskip 2mm {#1}}
\def\keywords#1{\vskip 1mm {\bf Kew words:} #1}
\def\reference{\vskip 6mm{\noindent{\bf References}} \vskip 2mm}

\def\r{\hangindent=1pc  \noindent}
\def\cen{\centerline}

\def\endpage{\vfil\break}

\def\kms{km s$^{-1}$}

\def\deg{$^\circ$}

\def\Vrot{V_{\rm rot}}

\def\kms{km s$^{-1}$}
\def\Msun{M_{\odot \hskip-5.2pt \bullet}}

\def\Deg{^\circ}
\def\deg{$^\circ$}

\def\halpha{H$\alpha$}
\def\ha{H$\alpha$}

\def\ioa{Institute of Astronomy, University of Tokyo, Mitaka, Tokyo 181, Japan}

\def\email{sofue@mtk.ioa.s.u-tokyo.ac.jp}


\cen{\bf The Most Completely Sampled Rotation Curves for Galaxies}

\vskip 6mm
\author{Yoshiaki SOFUE}

\inst{\ioa}

\inst{E-mail: \email}

\centerline{(To appear in ApJ)}

\vskip 8mm

\abstract{
We have compiled high-resolution position-velocity diagrams observed along the
major axes of nearby spiral galaxies in the CO-line emission, and
derived rotation curves for the inner regions of the galaxies.
We have combined the inner rotation curves with the outer HI and optical
rotation curves to obtain the total rotation curves.
The inner rotation curves are characterized by a steep increase
within a few hundred pc radius, indicating a compact massive concentration
near the nucleus.
We  fit the obtained rotation curves for individual galaxies
by a modified Miyamoto-Nagai's potential by assuming  existence of
four  mass components; a nuclear mass component with a scale radius
of 100-150 pc and a mass of $\sim 3-5\times 10^{9}\Msun$;
a central bulge of 0.5 to 1 kpc radius of a mass $\sim 10^{10}\Msun$;
a disk with scale radius 5 to 7 kpc and thickness 0.5 kpc of a mass
$\sim 1-2 \Mgal$;
and a massive halo of scale radius 15 to 20 kpc with a mass $\sim 2-3 \Mgal$.
We discuss the implication of the nuclear compact mass component
for the formation mechanism of multiple structures within the
galactic bulge during its formation.
}

\keywords{Galaxies: kinematics -- Galaxies: rotation --
Galaxies: structure --  Galaxies: general  -- ISM: CO emission --
ISM: HI gas}

\sect{1. Introduction}

The rotation curves of galaxies have been obtained  by optical (\halpha)
and  HI 21-cm line emission observations along the major axes
(Rubin et al 1980, 1982).
It is well known that the HI gas distribution generally shows
depression in the central few kpc region (Bosma 1981; Rots et al 1990),
which has yielded
an apparently solid rotation curve for the central  region.
Optical measurements are affected by the contamination of the
bright bulge light, which has also increased the uncertainty of the
curve near the center.
Moreover, because of the dust absorption in the gaseous disk,
optical observations cannot be obtained of the central regions
of  highly-tilted galaxies,
whereas edge-on galaxies are
most suitable for determining the rotation curves without
ambiguity of correction for inclination.

On the other hand, the CO-line emission is generally concentrated
in the central region, so that rotation curves of the inner few kpc
region can be most accurately  obtained
by CO position-velocity (PV) diagrams (Kenney \& Young 1988;
Young \& Scoville 1992; Sofue \& Nakai 1993; Sofue et al 1994).
It has been shown that the CO rotation curves of edge-on galaxies
do not necessarily coincide with those obtained by HI and/or optical
observations:
The central rotation shown
by CO is much flatter than that from HI and optical data,
or even increases near to the center, exhibiting rapidly rotating
compact disk component (Sofue et al 1988;
Sofue and Nakai 1993, 1994; Sofue et al 1994).

In this paper, we compile CO-line PV diagrams
along the major axes of nearby late type (Sb, Sc) galaxies
observed with large-aperture telescopes and interferometers.
Particularly, we extensively use CO-line data from the
Nobeyama 45-m telescope with an angular resolutions $15''$.
We then deduce inner rotation curves from the CO PV diagrams, and
combine them with outer HI and optical rotation curves,
and obtain the most completely sampled rotation curves.
Of course, this technique to deduce the inner rotation curve
works when nuclear CO is present:
significant quantities of nuclear CO might be not universal.
We further fit the obtained curves with the Miyamoto potential
(Miyamoto and Nagai 1975) by assuming four  mass components:
a nuclear compact mass, bulge,  disk, and a massive halo.

\sect{2. CO + HI  Position-Velocity Diagrams  and Rotation Curves}

{\subsection{2.1. Deriving Rotation Curves}}

\subsub{2.1.1. Edge-on galaxies}

Given a PV diagram along the major axis of
an edge-on galaxy, the rotation curve can be derived by
using the loci of terminal  velocity ($V_{\rm t}$)  in the PV diagram.
The terminal velocity is defined in a similar manner to that
adopted for HI and CO-line PV diagrams for our Galaxy (e.g., Clemens 1985).
Thereby, the velocity dispersion of the interstellar gas
($\sigma_{\rm ISM}$) and the velocity resolution of observations
($\sigma_{\rm obs}$) must be corrected  by
$$\Vrot=V_{\rm t}-(\sigma_{\rm obs}^2 + \sigma_{\rm ISM}^2)^{1/2}.\eqno(1)$$
For the galaxies discussed in this paper, the velocity resolution
was usually  $\sigma_{\rm obs}\sim 10$ \kms.
We take the interstellar velocity dispersion same as  that for molecular ISM
in our Galaxy,
$\sigma_{\rm ISM} \sim 7$ \kms\ (Stark and Brand 1989; Malhotra 1994).
In this paper, we adopted a correction for the
ISM velocity dispersion and  velocity resolution as
$\Vrot\simeq V_{\rm t}-12$ \kms\.
The accuracy of measuring  the terminal velocity as below using
PV diagrams was typically $\pm 10$ \kms.
So, the accuracy of determination of rotation velocities is not largely
dependent on the values of the velocity dispersion and resolution.

The terminal velocity is defined by a velocity at which
the intensity becomes equal to
$$I_{\rm t}=[(0.2 I_{\rm max})^2+I_{\rm lc}^2]^{1/2} \eqno(2)$$
on observed PV diagrams, where $I_{\rm max}$ and $I_{\rm lc}$ are
the maximum intensity and intensity corresponding to the lowest contour level,
respectively.
This equation defines a 20\% level of the intensity profile at a fixed
position, $I_{\rm t}\simeq 0.2 \times I_{\rm max}$,
if the signal-to-noise ratio is sufficiently high.
On the other hand, if the intensity is not strong enough, the equation gives
$I_{\rm t}\simeq I_{\rm lc}$ which approximately defines the loci along the
lowest contour level (usually $\sim 3 \times$ rms noise).

We comment that  thus-traced rotation curves may not
represent those corresponding to physically identical positions in galaxies,
since different galaxies have different CO/HI intensities, and as well,
their observational data are of different sensitivities and linear resolutions.
A threshold column density instead of the above criterion
might be a possible alternative.
However, it is practically hard to apply it to different galaxies with
different observational resolutions and sensitivities.
Hence, the practical, and probably most reliable way to derive the rotation
curves using radio PV diagrams is the method as described above.
Obviously, the result is observation-limited
(spatial and velocity resolutions and  sensitivity),
and depends on the proper line intensities of individual galaxies.

Fig. 1 shows an example of a  composite PV diagram for NGC 891
reproduced from Sofue et al (1994).
The CO gas is  concentrated in the central region,
while  HI is distributed in the outer disk, having
a void in the central region.
In Fig. 1 we superpose the thus obtained rotation curve for NGC 891 as an
example.
The HI gas indicates the rotation of the outer disk, whereas the CO emission
indicates the rotation in the innermost region including the rapidly rotating
nuclear disk.
The rotation curve as a function of the radius can be obtained by
averaging and smoothing the absolute rotation velocities in both sides of the
galaxy nucleus.
The final rotation curves are then smoothed according to the
angular resolution.
The clumpy and smaller-scale structures, which are partly due
to clumpy ISM distributions and molecular clouds and partly due to noise
in the observations, are also smoothed by hand in drawing
the final rotation curve.

In the following subsections, we derive rotation curves for individual
galaxies.
Generally, the rotation curves are almost flat even in the very inner
region, much flatter than those obtained from HI or optical observations.
We describe individual galaxies below, and summarize the observational
parameters and references in Table 1.

\Fig{1}

\cen{-- Table 1 --}

\subsub{2.1.2. Mildly tilted galaxies}

For nearly face-on galaxies that have been observed with a
sufficiently high angular resolution (e.g., sharper than a several tens pc),
this method will give an almost identical rotation curve
to that obtained by tracing intensity-weighted averages
(e.g., using a velocity field map), which
automatically account for the gas dispersion and velocity resolution.
However, except for such an ideal case,
both the finite beam and disk thickness
along the line of sight cause confusion of gases with
smaller velocities than the terminal velocity,
and would result in a lower rotation velocity.
Hence, even for galaxies with mildly-tilted galaxies
observed with a finite beam width, we use PV diagrams
along the major axes, and apply the same method as for edge-on galaxies.

\subsub{2.1.3. Innermost Rotation Curves}

This envelope-tracing technique has difficulty in applying to
the innermost part of the PV diagram, since simply traced envelopes on
the two sides of the nucleus have a discontinuity at the nucleus
mainly  due to the finite beam width.
We have avoided this discontinuity by stopping the tracing
at a radius corresponding to the telescope resolution, and then by
connecting the both sides of rotation curve by a straight
(solid-body like) line crossing the nucleus at zero velocity.
Therefore, the resolution of an obtained rotation curve
 is limited by the angular resolution of the observation.

The inner rotation curves are determined by CO data, while those
in the outer disk are determined from HI and optical data.
When we used data from different observations, we adopted
higher resolution data.
The data are then smoothly connected by tracing  higher-velocity parts.
Note that CO data have usually higher resolution than HI.
When comparable data were present in the same region, we simply averaged them.

\subsection{2.2. Rotation Curves for Individual Galaxies}

We present the thus obtained rotation curves for individual galaxies.
Basic parameters such as the distance and references
for individual galaxies are given in table 1.

\subsub{2.2.1. NGC 253}

For its proximity at a distance of 2.5 Mpc (Pence 1980), relatively
low-resolution CO data obtained with the FCRAO 14-m telescope (Scoville et
 al 1985) could resolve the central molecular disk at a linear resolution of
45$''$=545 pc.
A PV diagram obtained by Scoville et al (1985) indicates a steep increase
of the rotation velocity near the nucleus within $R \sim 0.3$ kpc.
An optical rotation curve has been obtained by Pence (1981), which
indicates a flat rotation at 2 to 5 kpc radius.
More outer rotation characteristics can be derived from an HI velocity
field observed by  Combes et al  (1977).

By combining rotation curves derived from these diagrams,
which are shown in Fig. 2a,
we constructed a total rotation curve as shown in Fig. 2b, where
the inclination of $i=78\Deg.5$ has been corrected.
Hereafter, figures a and b in each figure number will show
fitted curves to data and the resultant rotation curve, respectively.
The rotation velocity increases steeply in the central region,
and attains a maximum velocity of 210 \kms at $R\sim 0.3$ kpc.
Then, the curve is almost perfectly flat until $R\sim 9$ kpc

\Fig{2}

\subsub{2.2.2. IC 342}

This is an almost face-on ($i=25\Deg$) Sc galaxy at 3.9 Mpc distance.
It has been extensively studied in the CO line,
and various PV diagrams have
been obtained (Young and Scoville 1982; Hayashi et al 1987;
Sage and Solomon 1991).
Rotation curve in the HI line has been obtained by Rogstad and Shostak (1972).
We here make use of PV diagrams observed with the Nobeyama  45-m
telescope at a resolution of 15$''$ (284 pc; Hayashi et al 1987) and
a 4$''$-resolution mm-Array PV diagram (Ishizuki et al 1990a;
Ishizuki,  private communication).
Fig. 3 shows the obtained rotation curve for IC 342 using the PV diagrams.

The rotation velocity increases almost rigidly in the innermost region at
$R<10''=190$ pc, and reaches  $\Vrot\sim 130$ \kms at $R\sim 15'' (280)$ pc.
Then, it increases gradually to reach a maximum velocity at 190 \kms at
$R\sim 2-3'$ (2-3 kpc), followed by a flat HI rotation
at 195 (at 8 kpc) to 190 \kms (at 20 kpc).

\Fig{3}

\subsub{2.2.3. NGC 891}

This edge-on Sb galaxy at a distance of 8.9 Mpc has been extensively
observed in the CO line using the IRAM 30-m telescope
(Garcia-Burillo et al 1992), NRO 45-m (Sofue and Nakai 1993), and
OVRO interferometer (Scoville et al 1993).
curve.
It has been mapped in the HI line (Sancisi 1976a; Rupen 1991) at a
comparable resolution to the CO observations.
In Fig. 1 we show a composite PV diagram of the  CO and HI lines
reproduced from Sofue et al (1994).
The CO diagram is characterized by the 4-kpc molecular ring and
the high-velocity nuclear disk at $R<1$ kpc.
The HI gas is distributed in a broad ring and outskirts at $R>10$ kpc.
For deriving the rotation curve near the nucleus, we also made use of
the IRAM CO($J=2-1$) observation at a 13$''$ resolution (Garcia-Burillo
et al 1992) and the higher resolution PV diagram obtained by interferometer
observations by Scoville et al (1993).

The obtained rotation curve is given in Fig. 4.
After a steep rising up near the nucleus,
the rotation velocity attains a steep maximum
over 250 \kms, followed by a dip at $R=2$ kpc.
Then, it becomes almost flat at $R\sim 3$ kpc,
and remains so until  $R\sim 15$ kpc.
Beyond this radius, the rotation velocity gradually declines
toward the outermost region.
The rotation curve is very similar to that of our Galaxy (Fig. 11).

\Fig{4}

\subsub{2.2.4.  NGC 1808}

NGC 1808 is an Sbc galaxy known for its dusty jet
(V{\'e}ron-Cetty and V{\'e}ron (1985),
and the distance is  11.4 Mpc for a Hubble constant of 75 \kms Mpc$^{-1}$,
and the inclination angle is $i=58\Deg$.
HI observations using the VLA (Saikia et al. 1990) indicated a circular
rotation ring of about 7 kpc radius.
Koribalski et al. (1993) performed a mapping of the HI-line
absorption in the nuclear region using the VLA, and found a nuclear ring of
cold, dense rotating  gas disk of radius 500 pc.
Dahlem et al (1990) used the SEST 15-m telescope to map NGC 1808 in the CO
line emission at an angular resolution of $43''$, revealing a central
condensation of molecular gas.
We have mapped the central 1$'$ region  using the
Nobeyama 45-m telescope at a resolution of $15''$ in the CO emission,
and obtained a high-resolution PV diagram along the major axis
(Sofue et al: private communication).
This diagram shows a high-velocity rotating nuclear disk, consistent with
the result of Koribalski et al (1993).
These PV diagrams have been used to construct a rotation curve as shown
in Fig. 5.
The rotation speed increases steeply to 210 \kms at $R\sim 10$'' (500 pc)
in the nuclear region, and then decreases to 190 \kms at $R\sim 3$ kpc.
It increases again to a maximum of 210 \kms at $R\sim 2'$ (7 kpc) in the
HI rotation curve.
Beyond this radius, the rotation  declines to $\Vrot \sim 130$ \kms at
$R\sim 6'$ (18 kpc).
Such a declining  rotation in the outskirts is
rather exceptional among the galaxies studied here except M51 outskirts,
suggesting a small-mass massive halo.

\Fig{5}

\subsub{2.2.5. NGC 3079}

This is an amorphous edge-on galaxy classified as Sc type, showing an
anomalously high concentration of CO gas in the center (Sofue et al 1994).
The distance is taken to be 15.6 Mpc according  for the galacto-centric
HI systemic velocity and a Hubble constant of $H_0=$ 75 \kms Mpc$^{-1}$
(Sofue and Irwin 1992).
Fig. 6 shows the rotation curve produced by using the
composite CO + HI PV diagram obtained by Sofue et al (1994).
Here, they used a VLA HI PV diagram from Irwin and Seaquist (1991)
 and CO data from the Nobeyama mm Array (Sofue and Irwin 1992).
This galaxy exhibits an exceptionally high concentration of CO
emission in  the galactic center.
This high-density nuclear disk is clearly visible as the absorption
feature in the HI line.

The rotation velocity shows a steep rising-up to a maximum as high as
320 \kms in the SE side and 260 \kms in the NW,
followed by a dip at a few kpc radius.
The rotation velocity of this nuclear disk component is highly
asymmetric with respect to the nucleus.
The asymmetric rotation continues until $r\sim 8$ kpc.
The HI gas is widely distributed in
the broad ring at $R=1' - 2'$ (5 - 10 kpc) and in
the outskirts showing a symmetric flat rotation.

\Fig{6}

{\subsubsection{2.2.6. NGC 4565}}

This is an almost edge-on ($i\simeq 86\Deg$)
Sb galaxy at a distance of 10.2 Mpc.
A CO + HI composite PV diagram similar to Fig. 1 has been obtained
by Sofue et al (1994) who used CO data from  Nobeyama (Sofue and Nakai 1994)
and HI from the VLA (Rupen 1991).
The CO PV diagram shows a significant asymmetry in the intensity
distribution: the CO emission in the SE few kpc region is very weak, so
that the CO rotation in this is region is not clear.
However, except for this region, the total rotation characteristics is
almost symmetric, and mimics that of NGC 891.
On the other hand, the HI diagram shows an almost perfect symmetry
both in intensity and rotation velocity.

The rotation curve as obtained from these diagrams is shown in Fig. 7,
which is similar to that for NGC 891.
It has a nuclear-disk component rotating at 260 \kms,
followed by a flat rotation until 20-25 kpc at velocity as high
as $\sim 250$ \kms.
This galaxy is one of those with extremely flat rotation even in the
outskirts, suggesting a large extended massive halo.

\Fig{7}

{\subsubsection{2.2.7. NGC 5194 (M51)}}

This nearly face-on Sbc galaxy  at an inclination 20\deg and distance
9.6 Mpc  has been extensively studied in all wavelengths.
Tully (1974) derived a rotation curve from their optical spectroscopic
data for a wide area.
Rots et al (1990) have  extensively mapped this galaxy in HI, and
obtained an intensity-averaged HI velocity field.
However, they are not appropriate to derive an inner (a few kpc)
rotation curve, because the HI emission is very weak in the central
region (Rots 1990), so that the intensity-averaged velocity
is significantly weighted by rotation velocity at larger radius.
Unfortunately, no HI PV diagram has been obtained as yet along the major axis.
A high-resolution CO PV diagram has been obtained by Garcia-Burillo et al
(1993) and by Nakai et al (1995 in preparation).
We here use the CO PV diagram by Garcia-Burillo et al.
The outer rotation curve can be also obtained by the HI velocity field,
which agrees with that obtained from the CO data.

After correcting for the inclination of $i=20\Deg$,
we obtained CO and \ha\ rotation curves as shown in Fig. 8a.
The CO rotation velocity at $R<\sim 5$ kpc is
significantly higher than that from the \ha\ velocity.
This may be due to the fact that the density wave velocity jump
in the arms of M51 is  as high as $\sim 50$ \kms,
which would cause a systematic velocity difference of CO emitting regions
(dark lanes) from star forming regions (OB stellar arms) (Nakai et al
in preparation).
According to the definition of a rotation curve, we here simply
adopt the highest velocities (terminal velocities) along the major axis.
Hence, most part of the final rotation curve obtained
in Fig. 8b coincides with the CO rotation curve.
The rotation velocity increases steeply near the nucleus within 0.5 kpc,
reaching a maximum of 260 \kms.
Then, it remains flat up to 9 kpc, beyond which the rotation velocity
declines to 130 \kms at $R\sim 15$ kpc.
This declining rotation is similar to that observed in NGC 1808.

\Fig{8}

{\subsubsection{2.2.8. NGC 5907}}

NGC 5907 is a nearby Sc galaxy with an almost edge-on orientation at an
inclination angle of 88\deg.
Observations of the HI line emission have shown a large disk of interstellar
gas, which is warping in the outermost regions
(Sancisi 1976b).
A CO + HI composite PV diagram for  NGC 5907 has been obtained by
Sofue et al (1994), who used CO data from Nobeyama (Sofue 1994)
and HI data from the WSRT (Casertano 1983).
Recently, a nuclear disk component has been observed in the in CO $(J=2-1)$
line with the IRAM 30-m telescope, which showed two symmetrical humps in the
PV diagram at rotation velocities +200 and $-190$ \kms\ at 10$''$ to the NW
and SE of the nucleus.
This indicates that the rotation  velocity increases steeply in the
central $\sim 10''$ (500 pc).
Unfortunately, the HI PV diagram has been obtained only for the SE side,
which we used to derive an HI total rotation curve, assuming that
the rotation is axisymmetric.
So, we first obtained CO rotation curve at $R<10$ kpc by
averaging the SE and NW CO curves, and obtained final rotation curve
in Fig. 9b by smoothing the CO and HI total rotation curves.
After the steep rising near the nucleus, the rotation curve is almost flat
until 20 kpc, beyond which it is gradually declining.

\Fig{9}

{\subsubsection{2.2.9. NGC 6946}}

This is an nearly face-on galaxy  at a 5.5 Mpc distance,
and has been observed in high
resolution in the CO line (Sofue et al 1988; Ishizuki et al 1990b;
Casoli et al 1990).
Ishizuki et al (1990b) have obtained a CO PV diagram using the
Nobeyama mm Array at a resolution of 4$''$, which showed a very steep
rising of the rotation velocity up to a sharp maximum at 220 to 230 \kms
within the central 2$''$ (53 pc).
An inner rotation curve obtained from this PV diagram is shown in Fig. 10a-1.
Sofue et al (1988) have obtained a wide-area rotation curve by combining CO
data from the NRO 45-m observations with
an HI rotation curve by Tacconi and Young (1986).
They have shown that the rotation is almost perfectly flat from the very
center to the outskirts at $R\sim 15$ kpc.
Casoli et al (1990) have combined the CO PV diagram from IRAM 30-m
observations with an HI PV diagram, showing that
the rotation is almost flat toward the center.
We used this PV diagram to obtain a CO rotation curve shown in Fig. 10a-2.

We have combined all these rotation curves in Fig. a-1 and a-2,
 and obtained a rotation curve as shown
in Fig. 10a-3 and 10b.
Here we have corrected for  the inclination  of $i=30\Deg$.
The steep rising near the nucleus is followed by a decrease to a
dip at about 1 kpc, followed by a flat minimum at 185 \kms until
3 kpc.
Then, the rotation velocity gradually increases to attain 220 \kms at
$R\sim 7$ kpc, beyond which the rotation is nearly flat.
The flat rotation appears to continue until the observed edge of the
galaxy at $R\sim 16$ kpc.
The rotation curve  mimics the one for our Galaxy as shown in the
next subsection.

\Fig{10}

{\subsection{2.3. The Milky Way}}

The Milky Way Galaxy has long been observed both in HI and CO, and many
longitude-velocity ($l-V_{\rm lsr}$) diagrams have been published.
These diagrams have been used to obtain  rotation curves of the
Galaxy:
Burton and Gordon (1978) obtained an HI rotation curve at a
0\deg.5 resolution for the galactic disk within the solar circle,
and combined the HI data with CO data at $l>8\Deg$.
Clemens (1985) have analyzed the $^{12}$CO line survey data for $l>13\Deg$,
and integrated all the existence rotation curves
to present a total rotation curve of the Galaxy including the outer disk
of the solar circle.
In these studies, the rotation of the central few degree region
has been obtained by the terminal-velocity tracing method of the HI data
(Burton and Gordon 1978).
In Fig. 11 we reproduce the total rotation curve derived by Clemens et al
(1985).
Here, the solar rotation and radius are taken to be 220 \kms and 8.5 kpc,
respectively.

\Fig{11 a, b}

\subsection{2.4. Comparison of Rotation Curves}

All the rotation curves obtained in this work are shown in
the same scale in Fig. 12a and b.
Fig. 12a shows the rotation curves with a steep central peak, while
Fig. 12b shows those without.
Generally, the rotation velocity rises steeply within a few hundred
pc, indicating the existence of a central compact mass component.
Many  galaxies (the Milky Way, NGC 891, NGC 3079, NGC 6946)
exhibit a sharp maximum at $R\sim $ a few hundred pc,
reaching a velocity as high as $\sim 200$ to 300 \kms.
On the other hand, the maximum velocity corresponding to this component
is not so high in such  galaxies as NGC 253, IC 342, and  M51,
where the existence of the steep and sharp rising
near the nucleus is also evident.

\Fig{12}

\sect{3. Fitting by Miyamoto-Nagai Potential}

We try to fit the rotation curves by
the Miyamoto-Nagai (MN) (1975) potential.
The modified  Miyamoto and Nagai's (1975) potential with $n$
mass components is expressed in a $(R, z)$
coordinate as the following.
Here, $R$ denotes  the distance from the rotation axis and
$z$ is the height from the galactic plane.
$$
\Phi= \sum_{i=1}^n  {GM_i \over \sqrt{ R^2 + \left(a_i
+ \sqrt{z^2+b_i^2}\right)^2 }}, \eqno(3)
$$
where $M_i$, $a_i$ and $b_i$ are the mass, scale radius, and scale
thickness of the i-th mass component of the galaxy.
For a spherical mass distribution, we have $a_i=0$, and $b_i$
becomes equal to the scale radius of the sphere.
The rotation velocity is calculated by
$$
\Vrot = \left( R {\partial \Phi \over \partial R} \right)^{1/2}. \eqno (4)
$$

 Miyamoto and Nagai (1975) have assumed two components ($n=2$).
In order to fit the flat rotation at $R\sim 10 - 20$ kpc,
an extended massive hale has to be introduced.
Since their model has been proposed,
a three-component model ($n=3$) has been widely  used, which assumes
the central bulge, disk, and massive halo.
However, after a trial of fitting to the rotation curves of the central
few hundred pc region as obtained here, it turned out that the usual
three-component model is not sufficient to fit the steep central peak.
We have, therefore, introduced a fourth component which represents
a more compact nuclear component in addition to the usual three components

Fig. 13 shows an example of a calculated rotation curve of this
``four-component'' model ($n=4$), where we assumed
(1) a nuclear compact mass component, (2) bulge, (3) disk, and (4) a
massive halo.
The table inset in the figure presents the parameter combination.
Dashed lines indicate rotation curves corresponding to individual component.
The rotation curve of our Galaxy, except for the central 10-50 pc,
can be fitted by a model with
(1) a nuclear mass of $M_1=5\times 10^{9}\Msun$ of a $b_1=120$
pc scale radius;
(2) the bulge of $M_2=10^{10}\Msun$ and $b_2=750$ pc radius;
(3) the disk of $M_3=1.6\times 10^{11}\Msun$ with radius $a_3=6$ kpc and
thickness $b_3=0.5$ kpc;
and (4) a massive halo of $M_4=3\times 10^{11}\Msun$ and scale radius of
$a_4=b_4=15$ kpc.
The rotation of NGC 891 can be reproduced by the same model with
a similar parameter combination.

In this four-component model, however, the very inner rotation within
a few tens of pc region of the Galaxy, as shown in Fig. 11a-3,
cannot be reproduced.
Since it is beyond the scope of the present paper to discuss  a
detailed nuclear mass distribution,
we only argue for the necessity of introducing more central components.
In order to fit the observed inner rotation curve, we need to add a fifth
component at the nucleus with a smaller scale ($\sim 30$ pc
radius and mass ($\sim 10^7 \Msun$), and,
as well, a central point-like mass of a few $10^6 \Msun$.

\Fig{13}

Similarly, the rotation curve observed for NGC 6946 is fitted by the model
as shown in Fig. 14.
The flat valley at $R\sim$  1 to 2 kpc region can be fitted well by
introducing the four-component model, which was difficult to reproduce
by the three-component model.
The rotation of NGC 3079 can be fitted by a similar model.
The rotation curves of NGC 253 and IC 342 can be fitted by the same model
with a smaller-mass nuclear component, as shown in Fig. 15 and 16.
Rotation curves for the other galaxies can be also reproduced by this
model assuming parameters in between Fig. 13 to 16.

\Fig{14, 15, 16}

In the above fitting to the Miyamoto-Nagai potential with the
four mass components, we chose the parameters by trial and error.
Even through such a fitting, the parameters (mass and scale radii) can
be  constrained within an error of
about 10 to 20\%, depending on the quality and resolution of the data.
A detailed least-squares fitting to the data
would provide us of more realistic sets of parameters.

We have so far called the obtained diagrams the ``rotation curves''.
However, they actually meant observed loci of the
highest velocity envelopes in the position-velocity diagrams.
However, non-circular motion such as due to a barred potential and
density waves would be superposed on the actual motion of gas,
particularly in the central regions.
We, therefore, estimate the deviation of the tangential velocity
represented by the observed PV diagrams along the major axis
from that of a circular rotation.
Suppose that gas clouds are orbiting on elliptical orbits of eccentricity $e$.
Then, the orbital  velocity of a cloud at the perigalactic
passage is given by
$$V_0=\Vrot \sqrt{1+e}, \eqno(5)$$
where $\Vrot$ is the circular velocity corresponding to the mass
distribution as calculated by eq. (4).
The loci of maximum velocity on the PV diagram will approximate this
perigalactic (maximum) orbital velocity.

Therefore, the ``rotation curve'' may indicate a slightly
over-estimated circular velocity by a factor of $\sqrt{1+e}$.
For a highly disturbed orbits of gas in a strong bar shock
as numerical  simulations have shown (Fujimoto et al 1977;
 Huntley et al 1978; Noguchi 1988; Wada and Habe 1992),
the eccentricity is found to be of the order of $ e \sim 0.5$
Hence, the apparent rotation velocity from the PV diagrams would
be only $\sim 20$ \% higher than a purely circular velocity even
in such an extreme barred-shocked condition.
This would, however, result in an overestimation of the mass component
by a factor of $1+e$, causing an overestimation of a few tens of percent.
Finally, we mention that the rotation of the major gas
disk in the central 150 pc of our Galaxy has been
shown to be almost circular  from a detailed analysis of the CO PV
diagrams of Bally et al (1987) (Sofue 1995).

\sect{4. Discussion}

We have compiled position-velocity (PV) diagrams
along the major axes of nearby galaxies, which have
been observed in the CO line (central regions),
\Ha\ emission (star forming disk), and in HI lines (disk and outskirts).
We used these PV diagrams to obtain total rotation curves from
the nuclear region to outskirts.
The obtained total rotation curves are shown to be approximately flat
from the nuclear region of a few hundred pc radius to outer $R\sim 10-30$
kpc region, except for the inner few hundred pc.

A striking feature obtained in the present study is the
steeply rising nuclear peak of the rotation curves at
$R\sim 100$ to 200 pc, which is generally observed
for all the disk galaxies studied here.
This steep rotation peak can be fitted by a mass model in which
a compact nuclear mass component of a 100 to 150 pc radius and
a mass of several $10^9\Msun$ is assumed.
{}From a fitting of the observed rotation curves by  the
Miyamoto-Nagai (1975) potential, this nuclear mass component has turned
out to be an additional component to the well known central bulge:
The rotation curves of galaxies can be thus generally
fitted by a model with four mass components:
the nuclear compact mass, central bulge, disk, and the massive halo.

The nuclear mass component would have an essential implication
for the formation and evolution of the galactic bulge and the central
mass condensation of galaxies.
Saio and Yoshii (1990) have shown that the flat rotation curve and
exponential-raw mass distribution in disk galaxies are a consequence
of a viscous protogalactic disk contraction with on-going star formation,
where the time scales of viscosity and star formation are of the same
order, or of the order of the Jeans time of the disk instability.
Their model has also produced a central enhancement of the rotation
velocity at $R\sim 0.05 R_0$ with $R_0 $ being  the scale radius of the disk.
This is due to a more rapid contraction of the central gas
disk compared to the star formation time because of
a stronger shearing-viscosity in the central disk.
The model rotation curves could somehow  mimic even the central
velocity  peak of the observed curves such as in the Milky Way, NGC 891 and
NGC 6946.

However, the model appears to be still not satisfactory
in  reproducing in detail the steep central peak of rotation curves
at $R<\sim 200$ pc corresponding to the compact nuclear mass component.
In order for such a compact mass component to appear, a much more rapid
contraction of protogalactic gas disk would have been necessary.
Such a rapid contraction of gas disk prior to star formation
may be possible if we could modify (increase) the viscosity in the central
gas disk.
Alternatively, we may need to take into account a rapid gas accretion through
strong galactic shocks in a central oval (bar) potential (Noguchi 1988;
Wada and Habe 1992) during the proto-galactic disk contraction.

\reference

\r Bally, J., Stark, A. A., Wilson, R. W., and Henkel, C. 1987, ApJS 65, 13.

\r Bosma, A. 1981, AJ, 86, 1825

\r Burton, W. B., Gordon, M. A. 1978, AA 63, 7.


\r Clemens, D. P. 1985 ApJ 295, 422.

\r Casoli, F., Clausset, F., Viallefond, F., Combes, F., Boulanger, F.
1990, AA 233, 357.

\r Casertano, S. 1983, MNRAS, 203, 735.

\r Combes F 1992 ARAA, 29, 195.

\r Combes, F., Gottesman, S. T., Weliachew, L. 1977 AA 59, 181

\r Dahlem, M., Salto, S., Klein, U., Booth, R., Mebold, U., Wielebinski, R.,
Lesch, H. 1990, \aa, 240, 237.

\r{Fujimoto, M., and S$\phi$rensen, S. A., 1977, AA 60, 251}

\r Garcia-Burillo, S., Gu\'elin, M., Cerhicharo, J., Dahlem, M. 1992, AA
 266, 210.

\r Garcia-Burillo, A., Gu\'elin, M., Cernicharo, J. 1993, AA, 274, 123.

\r Garcia-Burillo, A., Gu\'elin, M., 1995 AA, in press 

\r Handa, T., Sofue, Y., Ikeuchi, S.,  Kawabe, R., and Ishizuki, S. 1992,
PASJ 44, L227.

\r Hayashi, M., Handa, T., Sofue, Y., Nakai, N., Hasegawa, T., Lord, S.,
Young, J. 1987 in Star Forming Regions, IAU Symp, No.115, eds. Peimbert and
J.Jugaku, D.Reidel Publ.Co. p.631.

\r Huntley, J. M., Sanders, R. H., and Roberts, W. W.,  1978, ApJ, 221, 521.

\r  Irwin, J. A., Seaquist, E. R. 1991 ApJ  371, 111.

\r{Ishizuki, S.  Kawabe, R., Ishiguro, M., Okumura, S. K., Morita, K. -I.,
Chikada, Y., and Kasuga, T. 1990a, Nature 344, 224. }

\r{Ishizuki, S., Kawabe, R., Ishiguro, M., Okumura, S. K., Morita, K. -I.,
Chikada, Y., Kasuga, T., and Doi, M.  1990b ApJ 355, 436}

\r Kenney, J., Young, S. J. 1988 ApJS 66, 261.

\r Koribalski, B., Dickey, J. M., Mebold, U. 1993, ApJL 402, L41.


\r Miyamoto, M., and Nagai, R. 1975, PASJ {\bf 27}, 533.

\r Malhotra, S. 1994 ApJ 433, 687.

\r{Noguchi, M. 1988, AA, {\bf 203}, 259.

\r Pence, W.D. 1980, ApJ, 239, 54

\r Rogstad, D. H., and Shostak, G. S. 1972, ApJ. 220, L37.

\r Rydbeck, etal 1985 AA 144, 282. 

\r Rots, A. H., Bosma, A., van der Hulst, J. M., Athanassoula, E.,
Crane, P. C. 1990, AJ, 100, 387. 

\r Rubin, V. C., Ford, W. K., Thonnard, N. 1980, ApJ, 238, 471 

\r Rubin, V. C., Ford, W. K., Thonnard, N. 1982, ApJ, 261, 439 

\r Rupen, M. P., 1991, AJ  102, 48.

\r Sage, L. J., Solomon, P. M. 1991 ApJ 380, 392.

\r Saikia, D. J., Unger, S. W., Pedlar, A., Yates, G. J., Axon, D. J.,
Wolstencroft, R. D., Taylor, K., Glydenkerne, K. 1990, MNRAS, 245, 397.

\r Saio, H., Yoshii, Y. 1990, ApJ 363, 40.

\r Sancisi, R. 1976a, AA  53, 159.

\r Sancisi, R. 1976b, in {\it Topics in Interstellar Matter}, ed. H. van
Woerden (D. Reidel Pub. Co., Dordrecht), p. 255.

\r Sandage, A., Tammann,  1974 ApJ 194, 559.

\r Sch\"oniger, F., Sofue, Y. 1993, AA, 283, 21.

\r Scoville, N. Z., Soifer, B. T., Neugebauer, G., Young, J. S., Mattheus, K.,
Yeka, J. 1985, ApJ., 289, 129.

\r Scoville, N. Z., Thakker, D., Carlstrom, J. E., Sargent, A. E.,
1993 ApJL 404 L63.

\r Sofue, Y. 1995, PASJ submitted.

\r Sofue, Y.  1994, PASJ 46, 173. 

\r Sofue, Y., Honma, M., Arimoto, N.  1994 AA in press 

\r Sofue, Y., Irwin, J. 1992, PASJ 44, 353.  

\r Sofue, Y., Nakai, N. 1993, PASJ 45, 139 

\r Sofue, Y., Nakai, N. 1994, PASJ 46, 147. 

\r Stark, A.A., Brand, J. 1989, ApJ 339, 763.

\r Tacconi,  and Young, J.  S.  1986, ApJ 308, 600.  

\r Tully, R. B. 1974, ApJ.S. 27, 437

\r Tully, R. B. 1988, Nearby Galaxy Catalogue (Cambridge Univ. Press).

\r{V{\'e}ron-Cetty, M. -P., V{\'e}ron, P. 1985, AA, { 145}, 425.}
[Outflow galaxy NGC 1808]

\r Wada, K., Habe, A. 1992, MNRAS 258, 82 .

\r Young, J S , Scoville, N Z 1982, ApJ. 258, 467.

\r Young, J S , Scoville, N Z 1992, ARAA, 29, 581.

\endpage
\settabs 9 \columns

\noi Table 1: Parameters for galaxies and references for PV
diagrams and rotation curves.
\vskip 2mm \hrule \vskip 2mm
\+ Galaxy & Type & Incl. & Dist.$^\dagger$ & Line & A. Reso. & L. Reso.&
References \cr
\+ & &(deg) & (Mpc) & &(arc sec) &(kpc) & \cr
\vskip 2mm \hrule \vskip 2mm
\+ N253 & Sc & 78.5 & 2.5 & CO &45 &0.55 & Scoville et al (1985) \cr
\+      &    &      &     &\Ha &-- &--&Pence (1981)    \cr
\+      &    &      &     &HI &120 &1.5&Combes et al (1977) \cr
\vskip 2mm

\+ IC342&Sc  &25    &3.9  &CO   & 15  &0.28  &Hayashi et al (1987)  \cr
\+      &    &      &     &CO & 4  &0.076 &Ishizuki et al (1990) \cr
\+      &    &      &     & CO   &45  &0.85  &Young\&Scoville (1982)  \cr
\+      &    &      &     & CO & 45  &0.85  &Sage and Solomon (1991)    \cr
\+      &    &      &     & HI & 50&0.9 &Rogstad\&Shostak (1972)  \cr
\vskip 2mm

\+ N891 &Sb  &88.3  &8.9  &CO &4& 0.17 &Scoville et al (1993) \cr
\+      &    &      &     &CO(2-1)&13 &0.55..&Garcia-Burillo et al (1992)\cr
\+      &    &      &     &CO & 15 &0.65 &Sofue\&Nakai (1993)    \cr
\+      &    &      &     &HI &20  &0.86 &Rupen (1991)    \cr

\vskip 2mm

\+ N1808 &Sbc &58   &11.4 &CO &15 &0.83 &Sofue et al (in prepa.)\cr
\+      &    &      &     &CO &45 &2.40&Dahlem et al (1990) \cr
\+      &    &      &     &HI &20 &1.1 &Saikia et al (1990) \cr
\+      &    &      &     &HI &20 &1.1 &Koribalski et al (1993)\cr

\vskip 2mm

\+ N3079&Sc  & $\sim90$&15.6 &CO &4 &0.30 &Sofue\&Irwin (1992)  \cr
\+      &    &      &     & CO &45 &3.4&Young et al (1988)\cr
\+      &    &      &     & HI &20 &0.15 &Irwin\&Seaquist (1991)\cr

\vskip 2mm

\+ N4565&Sb  &86    &10.2 & CO&15 &0.74 &Sofue\&Nakai (1993) \cr
\+      &    &      &     &HI & 20&1.0 &Rupen (1991)  \cr


\vskip 2mm

\+ N5194&~~Sc &20&9.6  &CO(2-1)&13 &0.60 &Garcia-Burillo et al(1993)\cr
\+ (M51)&    &      &     &\Ha &-- &-- &Tully (1974)  \cr
\+ &    &      &     &HI  &20 &1 &Rots et al (1990) \cr

\vskip 2mm

\+ N5907&Sc  & 88   &11.6 &CO&15 &0.84 &Sofue (1994) \cr
\+      &    &      &     &CO(2-1) &13 &0.7 &Garcia-Burillo, Guelin (1995)\cr
\+      &    &      &     &HI &40 &2.2 &Casertano (1983)    \cr

\vskip 2mm

\+ N6946&Sc  & 30   & 5.5 &CO &4 &0.11 & Ishizuki et al (1990) \cr
\+      &    &      &     &CO &15 &0.40 &Sofue et al (1988)  \cr
\+      &    &      &     &CO(2-1) &13 &0.35 & Casoli et al (1993) \cr
\+      &    &      &     &HI &20 &0.5 & Tacconi\&Young (1986) \cr
\+      &    &      &     & HI & 50&0.9 &Rogstad\&Shostak (1972)  \cr

\vskip 2mm

\+ Galaxy &Sb &90 &0  &CO, HI, HII &-- & --&Clemens (1985)  \cr
\+ Gal.Cen. &    &      &     &$^{13}$CO &2$'$ &5 pc &Bally et al (1987)    \cr
\+      &    &      &     & & & &Sofue (1995)    \cr

\vskip 2mm \hrule \vskip 2mm
$\dagger$ References to distances: NGC 253: Pence et al (1981);
IC 342: Tully (1988);
NGC  891: Handa et al (1992);
NGC 1808: Gal. cen. velocity 856 \kms, $H_0=75$ \kms Mpc$^{-1}$;
NGC 3079: Sofue and Irwin (1992);
NGC 4565: Sch\"oniger and Sofue (1993);
NGC 5194: Sandage and Tammann (1974);
NGC 5907: Sch\"oniger and Sofue (1993);
NGC 6946: Tully (1988).

\endpage

\noi Figure Captions
\v

Fig. 1: CO + HI composite position-velocity diagram for NGC 891
as reproduced from Sofue et al (1994).
A rotation curve is superposed by the thick line.

Fig. 2: (a-1) Inner rotation curve of NGC 253 derived by using the CO PV
diagram.
(a-2) CO + optical (\Ha) (= total) rotation curve compared to HI rotation.
(b) Total rotation curve of NGC 253 by averaging and smoothing the eastern and
western half of the rotation velocities.
See Table 1 for observational parameters and references.

Fig. 3: (a-1) CO rotation curves for IC 342 obtained by NRO 45-m telescope
and the  mm-wave array.
(a-2) CO rotation curves obtained by lower resolution observations
compared to those in (a).
(a-3) CO rotation curve obtained using curves in (a) and (b).
(b) Total rotation curve of IC 342 obtained by combining and smoothing the CO
curves with HI curve.

Fig. 4: (a) CO and HI rotation curves for NGC 891 shown separately
for the NE and SW part along the major axis.
(b) Total rotation curve of NGC 891.

Fig. 5: (a) CO and HI rotation curves for NGC 1808.
(b) Total rotation curve.

Fig. 6: (a) CO and HI absorption rotation curves for the inner region
combined with the HI rotation for the outer part.
(b) Total rotation curve for NGC 3079.

Fig. 7: (a) Inner CO and outer HI rotation curves for NGC 4565.
(b) Total rotation curve.

Fig. 8: (a) CO and optical rotation curves for NGC 5194 (M51).
(b) Total rotation curve of M51.

Fig. 9: (a) CO + HI rotation curves for NGC 5907.
(b) Total rotation curve.

Fig.  10: (a-1) CO rotation curve from the mm-array for the nuclear region
of NGC 6946.
(a-2) CO rotation curves.
(a-3) CO + HI rotation curves.
(b) Total rotation curve of NGC 6946.

Fig. 11:
Rotation curve of the Galaxy reproduced from Clemens (1985).

Fig. 12:  Rotation curves of galaxies studied in this paper
plotted in the same linear and velocity scales.

(a) Rotation curves having a central peak similar to that
of our Galaxy.

(b) Rotation curves without significant central peak.

Fig. 13: A model rotation curve of our Galaxy
as calculated for the Miyamoto-Nagai (1975)
potential with four components: (1) Nuclear mass component;
(2) Bulge component with scale radius
a few hundred pc; (3) Disk component; and (4) Massive halo component.
The mass $M$ in $10^{11}\Msun$,
scale radius $a$ and the thickness $b$ in kpc of each component are
indicated by the inset table.
The inset table shows the mass $M$ in $10^{11}\Msun$,
scale radius $a$ and thickness $b$ in kpc
for the four mass components of the modified Miyamoto-Nagai potential
in equation (3).
Dashed lines indicates rotation velocities corresponding to each mass
component.

Fig. 14: The same as Fig. 13, but mimicking that of NGC 6946.

Fig. 15: The same as Fig. 13, but mimicking that of
NGC 253 with a smaller-mass nuclear component.

Fig. 16: The same as Fig. 13, but mimicking that of IC 342.

\bye